\documentstyle [12pt,aasms]{article}

\begin{document}
\title{One-- vs. Two--Shock Heliosphere:  Constraining 
Models with GHRS L$\alpha$ Spectra toward $\alpha$ Cen
}
\author{K.G.\ Gayley, G.P.\ Zank, H.L.\ Pauls \\ Bartol Research Institute,
University of Delaware, Newark, DE 19716 \\
P.C.\ Frisch, and D.E.\ Welty \\Department of Astronomy and Astrophysics,
University of Chicago, IL 60637}
\date{October 4, 1996}

\begin{abstract}
Redshifted L$\alpha$ absorption toward $\alpha$ Cen
has been interpreted by Linsky \& Wood (1996)
and Frisch {\it et al.} (1996) as evidence for decelerated 
interstellar hydrogen piled up on the upstream side of the heliosphere. 
We utilize newly developed two-dimensional multi-fluid models of the
solar wind interaction with the ISM to corroborate this 
interpretation by synthesizing the L$\alpha$ absorption 
profile predicted for this ``hydrogen wall''.  
Both subsonic and supersonic inflow into the 
heliosphere are considered, corresponding to one-shock and two-shock 
global morphologies, respectively.  
It is found that these two extremes give observably different
redward absorption characteristics in the L$\alpha$ profiles, 
and our preliminary 
conclusion is that the L$\alpha$ profiles seen 
toward $\alpha$ Cen favor a barely subsonic model (Mach number 0.9). 
For such a model to hold,
additional interstellar pressure terms, such as cosmic ray or 
magnetic pressures, must contribute.
To make this conclusion more certain, 
an extended model-parameter survey is required, coupled with L$\alpha$ data
along other lines of sight.

{\it subject headings: interplanetary medium -- ISM: general -- shocks}
\end{abstract}

\section{Introduction}

The interaction of interstellar neutral particles with ambient plasma
near the boundary between heliospheric and interstellar material
(the heliopause) is of fundamental importance to the structure of
the heliosphere.
The length scale of the 
coupling between the neutral and plasma populations is on the order of
100 AU, which is much larger than the characteristic plasma scales
in the magnetic barrier at the heliopause.
This allows the neutrals to decouple from the
plasma, penetrate the barrier, and alter the heliospheric environment.
Indeed, excluding planetary atmospheres, $\sim$98\% of the diffuse gas
in the heliosphere originates in the surrounding interstellar cloud (SIC). 

The shock structure in the outer heliosphere consists of 
the termination shock (TS) where the
solar wind goes from supersonic to subsonic, the 
heliopause (HP), which is the stagnation surface between deflecting
interstellar ions and hot subsonic
solar wind plasma, and the possible bow shock (BS) where 
the interstellar flow goes from supersonic to subsonic.  
Heliospheric models with both a termination shock 
and bow shock are referred to as ``two-shock''
models, while models where the interstellar flow is subsonic are commonly 
referred to as ``one-shock'' models (no bow shock).   
The difference lies in the strength of the restoring forces 
that respond to plasma compression in the SIC.

The characteristic properties of the SIC have 
been determined from a combination of observations of nearby stars, 
observations of pick-up
ions in the solar
system, and direct observations of interstellar neutral helium at 5 au
(Piskunov {\it et al.} 1997; Witte {\it et al.} 1993).
The general cloud properties include neutral hydrogen density
$n_{tot} \sim $0.2 cm$^{-3}$, electron density $n_{e}$=0.1--0.3 cm$^{-3}$, 
and temperature $T \cong 7000$ K, which gives a plasma thermal 
sound speed of
$v_{th} \cong$ 14 km/s.
In the absence of cosmic-ray pressure, a bow shock forms 
around the heliosphere when the relative Sun-SIC velocity 
is larger than the magnetosonic velocity, which perpendicular to the
magnetic field is $v_{ms} = (v_{th}^{2} + v_{a}^{2})^{1/2}$, where
$v_{a} \sim 2.18 B(\mu$G$)/\sqrt{n_{e}} \ $ km/s
is the Alfven speed.
For characteristic
estimates 
$n_{e} \sim 0.1 $ cm$^{-3}$ and
$T\sim$7,000 K, and adopting
$B \sim 1.6 \mu$G (Frisch 1991),
the plasma thermal sound speed is 14 km/s and the Alfven speed
is 11 km/s, so the combined signal propagation speed is
$\sim$17 km/s.
Since the SIC has an observed velocity vector
of --26 km/s relative to the Sun
(arriving from the direction l=0$^{o}$, b=$+$16$^{o}$;
Witte et al. 1993; Bertin {\it et al.} 1993), this would imply
that the solar system has a Mach 1.5 bow shock.
For this reason,
two-shock models of the heliosphere (Baranov \& Malama 1993, 1995;
Pauls et al. 1995, Pauls \& Zank 1996a,b; 
Zank et al. 1996a,b; Williams et al. 1996a,b),
have received the most attention to date.

On the other hand, elementary estimates of the local
interstellar medium (LISM) parameters 
that include cosmic-ray pressure suggest that the interstellar flow may
be subsonic (Zank {\it et al.} 1996a,b).
Also, the possibility that the bow shock may be smoothed
out by ion-neutral effects in the presence of the interstellar magnetic
field, giving rise to an effective one-shock model, has been
suggested (Mullan \& Arge 1995).
Independent of these considerations, 
when interstellar neutrals are included self-consistently,
the neutral flow across the heliopause region is always seen
to be compressed, heated, and decelerated in the nose region
upstream of the heliopause.
However, since the detailed structure of the compressed
region, referred to as the ``hydrogen wall'', varies considerably between
one- and two-shock models, the availability
of sensitive diagnostics of this region would offer
a powerful probe of the global structure of
the heliosphere.

One potential diagnostic would be the direct observations of
such a wall in the heavily saturated L$\alpha$ line.
As will be shown in this paper, 
this pileup is detectable in directions where the
decelerated hydrogen is redshifted out of the shadow of the
interstellar absorption, if the interstellar column density is
low enough.
Linsky \& Wood (1996, hereafter
LW) and Frisch {\it et al.} 
(1996) attributed the redshifted excess absorption in
{\it HST} L$\alpha$ observations toward $\alpha$ Cen 
(earlier seen by {\it Copernicus} and {\it IUE},
Landsman {\it et al.} 1984) 
to the solar hydrogen wall.
Having galactic coordinates $l=316\deg \ $ and $b=-1\deg$,
$\alpha$ Cen lies $52{\deg} $ away from the upstream flow into the heliosphere.
This is close enough to the upwind direction to sample the hydrogen
wall, which should not be seen in downstream observations.
If the excess absorption is indeed of heliospheric origin, this
would provide a direct signature of the presence of the hydrogen wall
and a quantitative probe of its attributes.

In this paper, we present a direct comparison
between the $\alpha$ Cen L$\alpha$ data and the L$\alpha$ absorption predicted
by detailed heliospheric models that account for partially
coupled plasma/neutral hydrodynamics.
Our purpose is to test the following two key hypotheses, both of which will be
supported by our results.
(i) Synthetic L$\alpha$ absorption profiles generated from 2D models of
the solar-wind/LISM interaction provide strong theoretical support
that the observed redshifted absorption is heliospheric.
(ii) Quantifiably distinct L$\alpha$ absorption profiles arise from
one- and two-shock models, and a resolution of
$10^5$ (such as for the GHRS echelles) 
is sufficient to differentiate between them.

In \S 2 of this paper we summarize the observational evidence 
for the heliospheric H$^{o}$ pile-up toward $\alpha$ Cen.
The L$\alpha$ absorption from a homogeneous hydrogen wall
is considered in \S 3.
In \S 4, we introduce a more realistic hydrogen wall,
using multi-fluid heliosphere models 
developed by Pauls et al. (1995, 1996a,b),
Zank et al. (1996a,b), and Williams et al. (1996a,b). 
These model  predictions are compared with the 
$\alpha$ Cen spectra in \S 5.
In \S 6, we discuss the value of future observations along other sightlines,
and for each direction we predict the 
LISM column depth which would completely obscure the heliospheric absorption
signature for each of the models. 
In \S 7 we discuss how heliospheric models
can be used
to reduce uncertainties in observations of the upstream LISM D/H ratio.  
Our overall conclusions are summarized in \S 8.
Appendix A shows that the conductive interface models by Slavin (1989)
cannot reproduce the observations toward $\alpha$ Cen,
and in Appendix B we discuss our theoretical models in more detail.

\section{Signature of the Heliosphere in $\alpha$ Cen Spectrum}


%
Landsman et al. (1984) first discovered
that the centroid
of the foreground absorption seen in H$^{o}$ L$\alpha$ spectra of
the nearby solar twin $\alpha$ Cen A exhibited an
unexplained 8$\pm$2 km/s redshift with respect to the D$^{o}$ L$\alpha$ 
absorption feature.  
This conclusion was based on two {\it Copernicus } and eight 
{\it IUE}
spectra acquired over a period of  four years.  
They interpreted the velocity offset as evidence for two clouds
in this sightline.  
LW repeated the observation with the greatly improved
spectral resolution of the GHRS Echelle A, and
found evidence for at least two separate structures, one with the
expected LISM properties and the other more redshifted component with
a possible hydrogen pile-up at our heliosphere. 
They also explored the possibility that a third, {\it blueshifted} component
associated with the asterosphere of $\alpha$ Cen might be present in the data.
Other interpretations may be possible, but the overall redshift
of the H absorption could not be
created by any known interstellar cloud in the line of sight
(Lallement {\it et al.} 1995).
The possible presence of a heated
conductive interface along the line of sight is also an unlikely
interpretation, as argued in Appendix A.
Thus, the goal here is to critically explore the interpretation
that this redshifted feature is due to decelerated neutral hydrogen
at the heliosphere,
using detailed models which include recent advances
in heliospheric physics.

We first
show the LW observations in Figure 1, and describe our
approach for isolating the possible heliospheric signature.
%
%
The solid curves
in Figure 1 are the observed profiles for $\alpha$ Cen A (Figure 1a) and
$\alpha$ Cen B (Figure 1b).
The wavelength scale in both figures is relative to L$\alpha$ line center
in the heliocentric rest frame.
Note that the conversion to a velocity
scale is given by 0.1 {\AA } = 25 km/s.
The chromospheric emission lines show two obvious foreground
absorption features, a wide and saturated feature due primarily
to absorption by interstellar neutral H, and a narrow unsaturated
feature due to absorption by interstellar deuterium.

In order to determine the amount of absorption, the intrinsic stellar
emission profile must be specified.
We start with assumed intrinsic profiles shown as  
dashed lines in Figure 1.
The true profiles are not known for either star, although $\alpha$ Cen A
has the same spectral type (G2 V) as the Sun.
The secondary $\alpha$ Cen B is also a dwarf, though
is substantially cooler (K1 V).
Our approach is to take the intrinsic
solar L$\alpha$ profile
(Brekke {\it et al.} 1991) and rescale it linearly in both wavelength
and intensity to fit the wings of the observed profiles.
For the conclusions of our paper, an accurate
representation of the intrinsic stellar profile is not required,
since the features in which we are ultimately interested 
are quite sharp, representing absorption that varies dramatically
over a frequency interval that is comparable to the Doppler
width in the stellar L$\alpha$-forming region.
Any scheme for generating plausible profiles that vary only gradually over
such a narrow interval would be acceptable.

The next step is to
model the interstellar attenuation,
so that any residual absorption isolates the potential heliospheric signal. 
The dotted curves in Figure 1
give the attenuation of the stellar emission by an interstellar cloud with
neutral column depth
$N_H = 4.5 ~ \times ~ 10^{17}$ cm$^{-2}$, velocity 
$v=-18$ km/s, and Doppler broadening $b=9.3$ km/s.
These values are taken from the LW paper.
Here $N_H$ was fixed by scaling to the deuterium column density $N_D$ 
assuming D/H = $1.6 \times 10^{-5}$.
This D/H value is supported by downstream observations
(away from the heliospheric hydrogen wall) toward Capella (Linsky et al. 1993),
and is consistent with similar data in all directions (Wood et al. 1996). 
(In \S 6, we discuss the effects of modifying the assumed D/H ratio.)

Figure 1 shows clearly that
additional absorption both redward 
and blueward of the main interstellar feature is
required to complete the fit. 
Also, the additional absorption must be applied preferentially to
the redward side, so even if we arbitrarily increased the assumed ratio of
H to D, a fit could not be achieved.
LW and Frisch {\it et al.} (1996) interpreted the redshifted
absorption as a distinct component associated with the heliosphere.
Frisch {\it et al.} (1996) used early models by Zank {\it et al.} (1996b)
to model the heliospheric feature;
our goal here is to apply recent models that include an advanced
treatment of neutral/plasma coupling in the heliosphere for our
calculation of the expected L$\alpha$ absorption.

\section{Basic Diagnostics of a Hydrogen Wall}


L$\alpha$ absorption from neutral hydrogen that has been heated and 
decelerated in the region upstream of the heliopause provides
a useful probe of the physics in the heliospheric boundary layers.  
Prior to calculating the absorption signature
from detailed models, however, it is useful to 
explore in a general way
the effect of a neutral interstellar/heliosphere 
interface on incident L$\alpha$ profiles.
To do this, we consider the absorption for constant values of the
broadening speed $b_{hw}$ and velocity $v_{hw}$ over a specified 
column density $N_{hw}$. 
Since the column density of the 
hydrogen wall is expected to be $< 10^{15}$ cm$^{-2}$
(Zank {\it et al.} 1996), the
extended Lorentz wings of heliospheric hydrogen cannot
accumulate any appreciable opacity, so we restrict our 
analysis to the Doppler core.
 
The optical depth of the H$^{o}$ pile-up is then
\begin{equation}
\label{tau}
\tau_{hw} (\lambda) \ = \  {7.5 \times 10^{-13} \over b_{hw}} N_{hw}
e^{-(247 \lambda \ - \ v_{hw})^2/ b_{hw}^2} ,
\end{equation}
where $v_{hw}$ and $b_{hw}$ are in km/s
(and negative $v_{hw}$ corresponds to motion toward the
Sun), and $\lambda$ is in {\AA} from
line center in the heliocentric rest frame.
The narrow
absorption domain of interest in Figure 1 appears in the
vicinity of +0.1 \AA, corresponding
to a sub-population moving at +25 km/s away from the Sun.
LW found they could achieve a reasonable fit to the profile in 
this domain using $N_{hw} = 3 \times 10^{14}$ cm$^{-2}$,
$b_{hw} = 22$ km/s, and $v_{hw} = -8$ km/s, where we have averaged
their results for $\alpha$ Cen A and B when there were slight differences.
Using these parameters,
eq. (\ref{tau}) yields $\tau_{hw} = 1.12$ at $\lambda = +0.1$.
Thus, a simple constraint we can impose is that any
heliospheric model invoked to explain this
absorption feature must yield an optical depth of roughly unity 
at $\lambda = +0.1$ in the heliocentric frame.

Since the column depth of the hydrogen wall is three orders
of magnitude smaller than the column depth in the
LISM toward $\alpha$ Cen,
it may be surprising at first glance that the heliospheric optical
depth at +0.1 {\AA } is of the same order as the LISM optical depth
at that wavelength.
The key difference is that the neutral flow into
the hydrogen wall is heated and 
decelerated (and/or deflected), 
which both broadens and redshifts the heliospheric
component away from the -0.07 {\AA } centroid of the LISM absorption
and toward the +0.1 {\AA } wavelength of interest.

To compare the relative importance of the temperature increase and
the velocity shift in allowing the hydrogen wall to be visible,
we find simply from eq. (\ref{tau}) that decelerating the projected
velocity of the hydrogen wall along the $\alpha$ Cen sightline
by an additional 1 km/s (from -8 km/s to -7 km/s) has the same 
effect as increasing the temperature by 2300 K (if $b_{hw}$ is purely thermal,
so that $T_{hw} = 61 b_{hw}^2$).
Each would increase the optical depth by 15\%.
Since the LW $v_{hw}$ is redshifted by 10 km/s relative to the LISM,
and heated by about $24,000$ K,
crudely extrapolating the above analysis suggests that
each of these effects contributes about equally toward making the
hydrogen wall visible.
However, the nonlinear response to temperature rapidly
becomes important as the temperature falls, and eq. (\ref{tau})
indicates that $\tau(0.1)$ falls by a factor of 5 if $b_{hw}$ is
reduced to 16 km/s, corresponding to $T_{hw} \cong 16,000$ K.
For this reason, in the numerical results
below, the temperature is the parameter that shows the most significant
variations.
The velocity and column depth
structure also vary from model to model, however,
and they too affect the profiles.


\section{Model Characteristics}

Although heuristic models illustrate the overall hydrogen wall
characteristics,
the value of L$\alpha$ diagnostics is that they are sensitive
to the details of the heliospheric physics.
In this paper we apply such models, which include recent advances
in heliospheric simulation.
The LISM inflow parameters in the three models are summarized in Table 1,
and the details of the simulations are given in the above references.
Models 1 and 2 are drawn from previous work 
(Pauls {\it et al.} 1995; Zank {\it et al.} 1996a,b), and model 3
represents new results derived specifically to 
better fit the GHRS data.
For reference, we summarize the three models in Appendix B.

The hydrogen wall attributes depend on the assumed values of
the LISM 
inflow parameters (see Table 1), which include the density of electrons
($n_e$) and neutrals ($n_H$) in the surrounding LISM, their temperature $T$
(assumed equal), the speed ($v$) of heliospheric motion relative to the
LISM, and the ratio of $v$ to the propagation speed of
pressure disturbances in the plasma.
This last parameter, the Mach number, is critical for determining
the qualitative hydrodynamic response.
It depends on the pressure, which here is defined as
the thermal proton pressure times a correction factor $\alpha$,
\begin{equation}
P \ = \ \alpha n_p k T_p ,
\end{equation}
where $n_p$ is the proton density and $T_p$ is the proton temperature.

The parameter $\alpha$ (see Table 1) accounts not only 
for the electron pressure,
but also for any added contribution
from cosmic ray pressure or a perpendicular magnetic field.
Thus for a pure hydrogen plasma 
with $n_e = n_p$ and $T_e = T_p$, we have $\alpha = 2$ as in model 1.
The higher $\alpha$ values in models 2 and 3 imply additional contributions
which increase the effective sound speed, given by
\begin{equation}
v_s \ = \ \sqrt{\gamma \alpha k T_p/m_p} \ = \ 11.7 \sqrt{{\alpha T_p \over
10^4}} 
\end{equation}
measured in km/s, 
where $\gamma = 5/3$ is the adiabatic index and $m_p$ is the proton mass.
The increase in mean atomic mass
from helium and other species is not included, but would simply alter
$\alpha$ accordingly.

The plasma Mach number governs the qualitative
behavior of the heliosheath, and is thus
a key discriminant when considering the L$\alpha$
absorption.
The higher the Mach number, the greater the visible absorption,
due to the elevation of the temperature of the wall (cf. \S 3).
The heating of the plasma 
occurs not only due to adiabatic compression, but also to
charge exchange, which passes an electron from a neutral to
a proton. 
This important process
transports energy across the magnetic boundary between heliospheric
and LISM plasma, because
neutrals are not deflected by the magnetic field. 
They may therefore
cross the boundary freely and charge-exchange on the other side,
creating a neutral population with the attributes of the
plasma.
They may then return across the boundary and charge-exchange again,
and the net effect is to couple the partially
ionized plasmas across the magnetic barrier, as discussed in
more detail in Appendix B.

This process is 
especially significant for the heliosphere because the termination
shock strongly heats the solar-wind plasma, and charge-exchange 
with inflowing neutrals allows the
transfer of this heat into the plasma in the hydrogen wall region, which
in turn is transferred to the 
neutrals in the hydrogen wall via further charge exchanges.
The overall effect is to siphon off some of the solar-wind
kinetic energy flux and to deposit it as thermal energy, 
via three separate charge exchange
events, into the hydrogen wall.
Higher inflow Mach numbers allow this process to occur more
efficiently, by allowing a greater penetration of interstellar neutrals
into the termination-shock-heated plasma of solar-wind origin.
There is also more adiabatic heating for high Mach numbers, all of
which serves to elevate the characteristic $T_{hw}$.

As argued above, elevated temperature in the hydrogen wall is a key factor
in allowing the heliospheric L$\alpha$ absorption to be visible beyond the
wavelengths saturated by LISM absorption.
The other important factors were the deceleration
(or deflection) of the flow, and the column depth of the wall.
Thus for a simple comparison, in Table 2 we 
estimate a representative temperature $T_{hw}$, line-of-sight speed $v_{hw}$,
and column depth $N_{hw}$ of the hydrogen
wall along the 52\deg \ sightline
for each of the models in Table 1, and give for comparison the LW empirical fit.

Owing to gradients, the definition
of characteristic values is somewhat arbitrary.
We chose the values in Table 2 by averaging the
parameters at three separate points of interest 
in the wall.
These three points are where the temperature is maximal,
where the velocity is minimal, and where the contribution to
the optical depth at the key 0.1 \AA \ wavelength reaches its peak
(see Figure 4).
The value of the characteristic width $\Delta L_{hw}$
along the 52 \deg \ sightline is what would be required to
accumulate the same optical depth at 0.1 \AA \ as the actual
integrals over the model grid.
The equivalent column depth is then $N_{hw} = n_H \Delta L_{hw}$.
Also given to provide an overall length scale
is the distance to the heliopause
in the upstream direction $D_{hp}$, although the hydrogen wall
absorption is not highly sensitive to this model-dependent
parameter.

From Table 2 it is clear that higher Mach numbers yield greater heating
of the hydrogen wall.
If this conclusion is borne out by future parameter studies, then the
sensitivity of the width of the L$\alpha$ absorption to $T_{hw}$ gives
an observational constraint on the heliospheric Mach number.
In contrast, the characteristic line-of-sight velocity $v_{hw}$ does not vary 
appreciably for these models, while the equivalent column depth $N_{hw}$
varies as a complicated function of the inflow parameters.

It can also be seen that the simulation results are
substantially cooler than the heuristic LW fit, which would tend to
yield less L$\alpha$ absorption than is observed.
However, this is compensated by the higher amount of deceleration and/or
deflection in the simulations, i.e., a less-negative $v_{hw}$, which
creates a greater redshift of the heliospheric absorption relative
to the LISM.
This hints at the lack of uniqueness of observational fits, and shows
why the inclusion of heliospheric physics is essential.

Detailed depictions of the simulation results are shown in Figure 2.
The figure is described in detail in Appendix B.
Figures 2a and 2b give 2D plots of log $T$ contours, while the shading
shows the density distribution, for model 1 (two-shock) and model 2
(one-shock) respectively.
Figures 2c-e give line plots of the density, velocity, and temperature,
respectively, along the $\alpha$ Cen line of sight, where the dashed curve is
model 1, dotted is model 2, and dot-dashed is model 3.
Note the more gradual compression and deflection of the neutral flow in
the one-shock models, and the reduced peak temperature of the wall,
compared to the two-shock model 1.

\section{Lyman $\alpha$ Absorption Toward $\alpha$ Cen}


We are now ready to compare the detailed spectral features resulting from these
various models with the 
actual observations, thereby learning about the possible 
structure of our heliosphere.
The results are intended to be informative but not definitive, as a 
complete study over the full range of
plausible plasma and neutral inflow parameters, and
their application to multiple sightlines, is still needed.
This computationally demanding task is in progress. 
Here we demonstrate that the warm H$^{o}$ piled up against the
heliopause cannot be ignored when interpreting Lyman $\alpha$ data 
toward nearby stars in the upstream hemisphere.

The calculation of L$\alpha$ absorption by each model involves the
straightforward application of eq. (\ref{tau}), using the local
values at each gridpoint, and integrating  
over column depth along the $\alpha$ Cen sightline (52\deg \ from
the upstream direction).
The model grid extends to 1000 AU from the Sun, where the assumed
intrinsic profile corrected for LISM absorption 
over a Voigt profile is incident as a boundary condition.
The assumed LISM parameters, adapted from LW, 
are column depth $N_H = 4.5 \times 10^{17}$ cm$^{-2}$, 
Doppler broadening $b = 9.3$ km/s, and line-of-sight velocity $v = -18.2$ km/s.
The values of $b$ and $v$ are well constrained by observations
of other interstellar absorption lines toward $\alpha$ Cen,
and the effects of varying $N_H$ are considered in \S 7.

Figure 3 
shows the L$\alpha$ absorption at the red edge of
the LISM feature, for each of the 
heliospheric models listed in Table 1.
This can be compared directly with the GHRS data
from LW (solid curve).
The synthetic spectra are convolved with the instrumental broadening
of the GHRS Echelle A, which we take as a Gaussian of width 
0.008 \AA, though this correction is not essential for our purposes.
The Figure 3a/3b results refer to
the $\alpha$ Cen A/B data respectively.
Additional absorption at the blue edge is explained below.

The key discriminant of 
the H$^{o}$ pile-up is the quality of the fit at the red edge of the absorption
trough, between about 0.05 and 0.15 {\AA} from 
heliocentric line center, where the decelerated component is visible
outside of the saturated part of the interstellar line. 
The sloping absorption wings redward of this can be fitted with
relatively minor adjustments to the assumed intrinsic stellar profile,
so are not diagnostically significant.

Comparing the results of models 1--3 with the 
observations demonstrates the following
points, all of which are robustly insensitive to the uncertainties
in the intrinsic profile.
(i) Heliospheric L$\alpha$ absorption in
the supersonic model (model 1) is {\it too strong} due to the 
stronger deceleration and especially
the increase in temperature of the interstellar neutrals in the hydrogen wall. 
(ii) Heliospheric L$\alpha$ absorption in
the subsonic model with low Mach number (model 2)
is {\it too weak}, since the 
more gradually diverted interstellar plasma flow leads to less 
deceleration and less heating of the interstellar neutrals.
(iii) The model with a barely subsonic Mach
number of 0.9 (model 3) and a larger plasma density (see Table 1) 
does yield a favorable fit, giving compression
and charge-exchange heating of the neutrals intermediate between the
results of Models 1 and 2.

The consistency with GHRS data given by the parameters of model 3
should not be expected to be unique, and
other combinations could also suffice.
However, we strongly suspect that the incident interstellar gas flow 
can be neither highly supersonic nor highly subsonic, since these scenarios
lead rather inevitably to L$\alpha$ absorption that is either too strong or
too weak respectively. 
On the other hand, it appears that a barely subsonic 
interstellar wind provides the proper degree of both
deceleration and heating of the neutrals to fit the data.
The detailed constraints on the interstellar Mach number and inflow 
density imposed by
these data will be explored in future models.

Figure 3 also explores the cause of
excess absorption on the blue edge of the saturated LISM feature,
by following the LW speculation that it is due
at least partially to the joint asterosphere
of $\alpha$ Cen A and B.
At the level of a plausibility argument only,
we assumed that the $\alpha$ Cen asterosphere 
was identical to the heliosphere
(i.e., models 1, 2, 3 respectively), but viewed from the
appropriate angle (80\deg \ from the interstellar inflow direction).
This assumption is certainly not strictly warranted, but there are
some similarities in the two systems, since $\alpha$ Cen A is a
solar twin and the $\alpha$ Cen star system 
has a relative velocity with respect 
to its prevailing interstellar cloud of
22 km/s, similar to the Sun.

Note that absorption
by the solar hydrogen wall is seen only at the red edge of the LISM feature,
and similarly for the alpha Cen hydrogen wall at the blue edge, so that
the two sides of the absorption profile are linked only by the assumed
LISM attributes.
The schematic $\alpha$ Cen absorption at the
blue edge for models 1, 2, and 3 is
therefore also shown in Figure 3 with the same conventions
as the heliospheric absorption at the red edge,
and again offers promise for an acceptable fit.
This supports the view of LW that the $\alpha$ Cen asterosphere
is detected, and offers exciting possibilities for
modeling the wind/LISM interaction in this system.
Of course, alternative explanations are possible; for example, 
we show in Appendix A that a cloud interface could mimic 
a stellar HW for this sightline.

To understand the integrated absorption in terms
of its spatial contributions from the local neutral
parameters over the grid,
Figure 4
shows the contribution to the optical depth at 0.1 \AA \ 
compiled over 100 AU intervals.  
Since the area under the curves gives the integrated
optical depth at the key wavelength 0.1 \AA \, 
it provides a proxy for the overall
impact of the heliosphere on the observed L$\alpha$ profiles.
The importance of H$^{o}$
heating in producing observable hydrogen-wall absorption is evident since
the region of greatest temperature (Figure 2e) 
correlates with the region of
maximum contribution to the optical depth.

\section{Heliospheric Contributions in other Sightlines}

Observations along the single $\alpha$ Cen sightline yields 
valuable yet limited information
about the inferred Mach number of solar motion through the LISM.
Data from sightlines that cross the hydrogen wall along other
angles would be extremely
helpful for confirming the heliospheric origin of the absorption, and
would provide additional constraints on the heliospheric parameters.
LISM column depths substantially above
$10^{18}$ cm$^{-2}$ completely blanket the wavelength domain
where heliospheric absorption could have appreciable opacity, therefore
useful L$\alpha$ sources are scarce.
Even if some starlight penetrates at wavelengths of heliospheric
absorption, the bandwidth of the signal may be too narrow to
be spectrally resolved.
The purpose of this section is to derive the equivalent width
of the L$\alpha$ heliospheric absorption for a given model, 
to determine the usefulness of a given stellar source
as a heliospheric probe, as a function of LISM column
depth and instrumental resolution.

The estimate is obtained by assuming a flat stellar spectrum incident
on the LISM cloud with values $b=9.3$ and $v = -29$ cos $\theta$ km/s
(taken from LW),
where $\theta$ is the angle from the upstream direction, and the
column depth $N_H$ is treated as a variable.
The stellar spectrum transmitted through the LISM, as a function of $N_H$, 
then provides the input for the calculation of the heliospheric
absorption feature, 
for which we compute an equivalent width as a function of $N_H$.
If the absorption is significant at any wavelength,
this equivalent width gives a conservative estimate of
the wavelength resolution required to clearly distinguish the feature.

The equivalent widths of the
model heliospheric absorption are plotted as contours in Figure 5,
as a function of $N_H$ for each sightline,
for heliospheric models 1 and 2 (shown in Figure 5a and 5b respectively).
Model 3 is intermediate to these results.
The value of $N_H$ that corresponds to each point on a given contour
is the radial distance from the origin, measured by the scale
given on the abscissa and ordinate.
The angle of the sightline from the upstream direction
maps directly into the angle from the abscissa, so the figure
represents orientations and column depths in real space with the
heliosphere at the origin.
The sightlines to 
36 Oph, $\alpha$ Cen,
and 31 Com would be at $\theta=$ 12\deg, 52\deg, and 72\deg \ respectively,
and are indicated.
The absorption equivalent widths are measured in \AA.
It is expected that observed resolution elements somewhat larger than
the absorption equivalent widths 
in the figure may still be valuable if the signal-to-noise
is high, so the contours merely provide guidelines for the
preferred degree of resolution to guarantee useful diagnostics.

As an example, the resolution of the GHRS Echelle A 
corresponds to a Gaussian Doppler width of
$\sim$0.008 \AA \ at L$\alpha$.  
According to Figure 5, this means that along the sightline to $\alpha$
Cen, the heliospheric
absorption can be resolved easily for LISM columns below about 
$0.7 \times 10^{18}$ cm$^{-2}$
and 
$1.2 \times 10^{18}$ cm$^{-2}$
for the one-shock and two-shock models respectively.
As another example,
by interpolating for model 3 along the sightline to 31 Com we conclude that
the heliosphere should be easily resolvable to the GHRS echelle for 
intervening column densities of N$\le$$1 \times 10^{18}$ cm$^{-2}$.
Since the interstellar $N_H$ column toward 31 Com is indeed around 
$1 \times 10^{18}$ cm$^{-2}$ (Piskunov, Wood, \& Linsky 1996),
this illustrates the potential value of searching for heliospheric 
H$^{o}$ absorption in high-resolution L$\alpha$ observations of stars other
than $\alpha$ Cen. 
At the time of this writing,
an unpublished analysis of the L$\alpha$ absorption toward 31 Com
by Dring {\it et al.} (1997) was unable to identify a clearly heliospheric
absorption component.
The quantitative constraints this imposes on the heliosphere are still
being examined, but will presumably provide further evidence against
strongly supersonic plasma inflow.

%
%

\section{The D/H Ratio}

%
%

Since our fundamental constraint 
for distinguishing between one-shock and two-shock
models is the degree of excess absorption illuminated by
starlight transmitted by the LISM, 
the LISM hydrogen column is crucial to specify.
This is done by scaling to the easily inferred deuterium column, assuming
that the D/H ratio takes the value that is canonical in the
solar neighborhood along 
various other sightlines (Piskunov, Wood, \& Linsky 1996), which 
is $\sim 1.6 \times 10^{-5}$ (by particle, not mass).

However, if this ratio varies in the LISM, and were to be
lower along the $\alpha$ Cen 
sightline, then the additional LISM hydrogen absorption 
would require less heliospheric
absorption, favoring models with an even lower Mach number.  
Indeed, if D/H were to be so low
as a factor $\sim$3, then the
{\it one-shock} model 2 results fit both edges of the absorption extremely well,
and no contribution from absorption by the $\alpha$ Cen hydrogen wall
would be required toward the blue.
Contrarily, {\it increasing} D/H by an order of magnitude, however 
implausible, would move the {\it two-shock} model 1 results
into better agreement, and the $\alpha$ Cen wall would then also
have to contribute more strongly.
This implies a connection between the assumed D/H ratio and the best-fit
heliospheric model.
Therefore, a variety of heliospheric models could be consistent with 
the data, but only if the D/H ratio can be varied by about an order
of magnitude. 

Since our conclusions are insensitive to small 
variations in D/H, it is apparent
that constraint information travels more easily in the direction from 
knowledge about D/H into knowledge about 
the heliosphere, rather than the converse.
With the inclusion of future observations along other lines of sight
through the hydrogen wall, a suitable
synthesis between the observational ramifications of both D/H and heliospheric
absorption can be achieved,
and confidence in a stable
upstream D/H ratio may be further established. 
In light of the cosmological significance of D/H, this represents a
rather unique interplay between heliospheric
and astronomical areas of interest.

\section{Conclusions}

We have shown that high spectral resolution observations of L$\alpha$
absorption, coupled with careful modeling of the interaction of the
solar wind with the LISM, provide a useful diagnostic for remotely
sampling the global structure of the heliosphere.  
Our conclusions are as follows.

(i) Several models (Baranov \& Malama 1993;
Pauls {\it et al.} 1995; Zank {\it et al.} 1996a,b)
have independently corroborated the theoretical expectation that
warmed interstellar neutrals should
accumulate upstream of the heliopause.
Our synthetic absorption profiles for three distinct heliospheric models
support the detection of this hydrogen wall
toward $\alpha$ Cen by Landsman {\it et al.} (1984)
and LW, and our barely subsonic
model (Mach number 0.9) closely resembles the observations.
We conclude that the hydrogen wall upstream of our heliosphere has
indeed been seen.
A particularly promising avenue for concretely confirming this conclusion
is to combine the $\alpha$ Cen data with archival high-resolution 
L$\alpha$ spectra of 31 Com, since this would provide stereoscopic
sampling of the heliosphere at angles $52\deg$ and $73\deg$ from the 
upstream direction.

(ii) Our model results indicate that the differences between a one- and
two-shock heliosphere manifest themselves in observably distinct
L$\alpha$ absorption features, providing a powerful discriminant
between these possibilities.
Furthermore, application of this diagnostic 
exerts quantifiable constraints on the plasma and
neutral environment in the inaccessible LISM.
For example, the LISM 
parameters (temperature, density, flow speed, and fractionation)
from model 3, which yield a reasonable fit to
the L$\alpha$ data, are
only in good agreement with existing estimates (Frisch 1995, Gloeckler
1996) if we further stipulate that cosmic rays
(or some other mechanism, such as a perpendicular magnetic field) contribute
appreciably to the total interstellar pressure.
The possibility that cosmic rays (e.g., Holzer 1979) or magnetic fields
(e.g., Mullan \& Arge 1996) could serve
in this way to reduce the inflow Mach number
is not unexpected, but since
we have yet to undertake a comprehensive study over the full range of
possible LISM inflow parameters, we cannot at present establish the
uniqueness of the Mach 0.9 fit.
Complementary observations, such as backscattered solar L$\alpha$
from neutrals in the heliosphere, may provide 
invaluable additional constraints,
since the FWHM is sensitive to the neutral heating.
Thus although our results are 
highly suggestive, conclusive evidence of subsonic inflow awaits future work.

(iii) Models that include the charge-exchange coupling between
plasma and neutrals have established the importance of a process,
described in detail by Zank {\it et al.} (1996a) and Zank \& Pauls (1996),
whereby interstellar neutrals that charge exchange with very hot
shocked solar wind plasma (i.e., the ``component 2'' neutrals described
in Appendix B)
subsequently stream out into the LISM.
Although of low density 
($\sim 10^{-4}$ cm$^{-3}$), these neutrals are very hot ($\sim 10^6$ K), 
and after a second charge-exchange process,
deposit considerable heat into the plasma just outside the heliopause.
This temperature increase acts to further heat
the hydrogen wall via subsequent charge exchanges.
The potent combination of heating and deceleration produces visible
absorption in the L$\alpha$ spectra from $\alpha$ Cen, and this
absorption becomes more pronounced as the peak temperature increases.
Models which neglect heat transport across the heliopause 
via charge exchange cannot produce the
proper L$\alpha$ opacity profile.
Furthermore, the evidence that hydrogen-wall absorption is indeed
clearly visible implies that some such heating mechanism must
indeed be operating, which supports this hitherto purely
theoretical picture.

(iv) Although the blueshifted absorption excess has not received close 
attention here, it was found that if $\alpha$ Cen has
a relatively solar-like asterosphere, then its hydrogen wall should also
be visible.
This supports the suggested detection of an $\alpha$ Cen hydrogen
wall by LW,
and opens new possibilities for constraining models of
the wind/LISM interaction in this nearby stellar system.

GPZ and HLP were supported in part by NASA grants 
NAGW-2076, NAGW-3450, NSF Young Investigator Award ATM-9357861, and a Jet
Propulsion Laboratory contract 959167. 
KGG acknowledges the support of NSF grant
AST-9417090; PCF acknowledges the support of NAGW-5061, NAGW-2610 and NAG5-3176;
and DEW acknowledges the support of NAG5-3228.
We are indebted to Dr. Brian Wood for providing the 
GHRS $\alpha$ Cen data and for helpful input on the text, and appreciate
the cooperation of Dr. Jeffrey Linsky in this project.  
We also want to thank Dr. Jon Slavin for
providing the opacity profile for warm neutral hydrogen expected 
in the interface between the local interstellar cloud and surrounding
hot plasma.  
The computations were performed in 
part on the CRAY-YMP at the San Diego Supercomputer Center. 

\newpage
\section{Appendix A}
 
In this Appendix, we discuss other potential sources of
excess L$\alpha$ absorption that could mimic a heliospheric feature, and
argue that, although other possibilities cannot be
competely ruled out, none provide as natural and 
theoretically straightforward an explanation as the hydrogen wall.
Since Landsman et al. (1984)
concluded that the observed $\sim +8$ km/s offset between the
centroids of the H$^o$ absorption and the D$^o$
absorption could not be explained by uncertainties in the underlying 
stellar emission profile, and
Linsky \& Wood (1996b) ruled out uncertainties due to geocoronal emission, 
we must conclude that the foreground
L$\alpha$ absorption is intrinsically complex. 

Since the H$^o$ L$\alpha$
optical depth of the LISM feature lies on the saturated
part of the ``curve of growth'', whereas the LISM D$^o$ 
and the allegedly heliospheric H$^o$ features lie in
the unsaturated linear domain,
the centroid of the $H^o$ absorption could be sufficiently
shifted relative
to the D$^o$ if there exists a subset of material that is optically thick in
H$^o$ but not in D$^o$.
Since this roughly 8 km/s offset
is caused by H$^o$ absorption 
near 0.1 \AA \ (+25 km/s relative to the LISM), such small column
depths could only be effective if a component exists that has
a bulk velocity shift of this magnitude,
or if the velocity dispersion approaches 25 km/s within a component
that is itself shifted by at least 8 km/s,
or some combination of the two.

Such a dispersion cannot be accomplished by overlapping
pieces of observed LISM bulk flows, since
the velocity difference between the main component
toward $\alpha$ Cen and the SIC (also called the LIC)
clouds is only $\sim 2.5$ km/s in this line of sight
(Lallement {\it et al.} 1995).
Therefore, we must either assume that this velocity
dispersion is achieved at the atomic level via warming of a small
fraction of the foreground material to
temperatures of at least about 20,000 K, or postulate the
existence of thin ``wisps'' moving
at speeds around 25 km/s relative to the bulk of the LISM.
There is no evidence in favor of the existence of
fast-moving wisps, 
so 
we focus on possible absorption due to 
the warm interfaces speculated on more physical
grounds to exist (Slavin 1989) between 
bulk LISM clouds and the hot ISM plasma filling nearby space.
Note that a cold component at the SIC velocity would not yield
sufficient opacity at the wavelengths of interest to affect our conclusions.

In the model of
Slavin (1989), the LISM interface is
a $\sim$0.0032 pc layer of H$^{o}$ with 
N(H$^{o}$)$\sim$3.5 $\times$ 10$^{14}$ cm$^{-2}$, 
$T \sim$ 8,000--80,000 K, 
and $v \sim$ 1 km/s with respect to the cloud.
%
We found that the theoretical velocity distribution of the heated interface
neutral H yielded absorption that could be most simiply fit by the sum of the
two Voigt absorption features;
on with 
and the other with N(H$^{o}$)$\sim$1.0 $\times$ 10$^{14}$ cm$^{-2}$
and $b=28$ km/s, both moving at 0.9 km/s with respect to the underlying cloud
from which it is evaporating.
Our test model assumes that two such interfaces exist toward $\alpha$ Cen, 
one at the edge of the main
cloud toward $\alpha$ Cen (-19.1 km/s),
and the other at the edge of the 
local (SIC) cloud (at -15.5 km/s).
The thermal broadening and total column depth of 
a warm Slavin interface for each cloud are similar to the heliopause models,
and potentially difficult to distinguish.
However, such interfaces do not yield sufficient velocity
offset between the observed H$^o$ and $D^o$ features.
We added two hypothetical interfaces to the LW LISM absorption
without any heliospheric component, and obtain the
results shown in Figure 6.

We conclude that the presence of Slavin interfaces would 
affect observed LISM L$\alpha$ absorption and could complicate
the interpretation of the heliospheric signal, although a heliospheric
contribution would still be required to yield the observed redshift.
Thus, the possibility that the heliospheric signal could be blended with
warm ISM structures remains open, but there is presently no
consistent evidence that this is the case.  
%
%

\newpage

\section{Appendix B: 
Multi-fluid Heliosphere Models with Sub- or Supersonic Inflow}

Two groups have developed detailed models of the interaction 
between the solar wind
and the LISM that include the neutral hydrogen component
self-consistently with the plasma.
A two-dimensional (2D) Monte Carlo approach has been
pioneered by Baranov and Malama (1993, 1995), while 
Pauls et al. (1995, 1996a,b),
Zank et al. (1996a,b), and Williams et al. (1996a,b,
all summarized in Zank \& Pauls 1996), have instead chosen a
multi-fluid approach to describe the neutral hydrogen.
Both yield
similar results for identical input parameters, although the simulations
of the latter group have recently
been extended to three dimensions (Pauls \& Zank, 1996b).
These latter models are also time dependent, but approach a nearly
steady state.
Since the details are described in the
above references, we present here only an overview of
the salient features of the 2D multi-fluid models we utilize
for our analysis of the L$\alpha$ absorption.
 
The key physical process responsible for altering the dynamics of
interstellar H$^{o}$ at the heliopause is the coupling
to the local plasma via charge exchange, whereby 
the bound electron from a neutral is passed to a proton 
during collision without any other effect, thus swapping the neutral and
proton attributes.
To avoid solving the neutral Boltzmann equation directly to describe
this interaction, we recognize that, to a good
approximation, there exist essentially three distinct neutral H components
(Holzer 1972; Hall 1993) corresponding to three physically distinct
regions of origin.  
Neutral H atoms whose source lies beyond the heliosphere
(region 1) are component 1. 
This ``thermal'' neutral 
component is thus interstellar in origin,
although dynamical changes in the distribution result from 
charge exchange with interstellar plasma which has been
significantly affected by the presence of the heliosphere. 
By contrast, neutral component 2, which is born
via charge exchange in the solar-wind
shock-heated heliosheath and heliotail
(region 2), is suprathermal compared to component 1, its high
temperature reflecting that of the local $10^6$ K plasma.
Although of low density, 
component 2 can be dynamically important owing to its ability
to transport heat across the heliopause.  
The third (``splash'') component is produced in
the cold supersonic solar wind prior to reaching the
termination shock (region 3), and this
very tenuous component is characterized by high radially outward velocities.

Each of these three neutral components is represented 
by a distinct Maxwellian distribution function appropriate to the
characteristics of the source distribution. 
This then allows us to simplify the
production and loss terms corresponding to each neutral component
(Ripken \& Fahr 1973; Zank {\it et al.} 1996a; Williams {\it et al.} 1996a).
The complete highly non-Maxwellian H distribution 
function is then the sum over the three components, i.e.,
\begin{equation}
\label{eq:1}
f ({\bf x} ,{\bf v} ,t) = \sum_{i=1}^3 f_i ({\bf x} ,{\bf v} ,t) . 
\end{equation}                     
Equation (\ref{eq:1}) allows us to express the full Boltzmann equation for the
neutrals as distinct equations corresponding to each of components 1, 2 and 3,
all coupled through their respective production and loss terms.
From each of these component Boltzmann equations, one
obtains a distinct isotropic hydrodynamic description for each of the three 
neutral components. 
Thus, our multi-fluid description comprises all three 
neutral fluids coupled to a hydrodynamic plasma. 
This rather computationally 
demanding time-dependent system of equations is solved in two spatial 
dimensions using a method developed by Pauls et al. (1995).

\subsection{Two-Shock Model}

Here we summarize the 2D time-dependent two-shock results obtained using the 
multi-fluid model.
For a complete description, see
Zank {\it et al.} (1996a) and Williams {\it et al.} (1996b). 
Table 1 lists the parameters (model 1) used for the simulations. 
In Figure 2a, 
a 2D plot of the component 1 neutrals ({ i.e.,} neutrals
of interstellar origin) is presented. 
The contours denote
Log($T$), the arrows show the flow direction and the shading describes
the density normalized to the inflowing interstellar gas density. 
The hydrogen wall, where the inflowing H$^{o}$ is compressed 
between the BS and HP in the upstream direction, is apparent. 
As described above, Figure 2c-e 
shows the 
$\alpha$ Cen line-of-sight profiles for the density, 
velocity and temperature, respectively, of component 1, which
is the source of the observable absorption
(components 2 and 3 are too rarified to yield significant features).
A detailed presentation of the plasma
and component 2 and 3 neutral results is found in Zank {\it et al.} (1996a). 
The dashed curves denote the results for the two-shock case
(model 1, Table 1), whose key features can be summarized as follows. 
\begin{description}
\item[(i)] The TS is located at $\sim 95$ AU, the HP at $\sim 140$ AU and the BS
at $\sim 310$ AU in the upstream direction. In the sidestream and downstream 
directions, the TS is located at $\sim 140$ AU and $\sim 190$ AU respectively. 

\item[(ii)] Inflowing component 1 neutrals
are decelerated substantially and filtered
by charge exchange with the interstellar plasma between the BS and HP in 
the upstream direction, 
which leads to the formation of
a hydrogen wall with densities up to about $0.3$ cm$^{-3}$, column depths
up to about
$10^{14}$ cm${}^{-2}$, and temperatures ranging from 20,000 K to 30,000 K. 
The pile-up in the neutral gas results from the deceleration and 
deflection of the neutral flow by 
charge exchange with the interstellar plasma,
which is itself decelerated and diverted due to the presence of
the heliosphere. 

\item[(iii)] Component 2, produced via charge exchange between component
1 and hot shocked solar wind plasma between the TS and HP, leaks
across the HP into the cooler shocked interstellar gas and 
heats the plasma through a {\it second} charge exchange. 
This leads to an extended thermal foot abutting the outside edge
of the HP. This heating of the plasma by component 2 serves to broaden the
region between the BS and HP, as well as to (indirectly) further
heat the component 1 interstellar neutrals
after {\it subsequent} charge exhanges. 
Some minor heating of the unshocked LISM also
occurs upstream of the BS, thereby
marginally reducing the Mach number of the incident interstellar wind. 

\item[(iv)] The temperature of component 1 neutrals once inside
the heliosphere remains fairly constant
in the upstream region, at $T \sim$ 20,000 K, a substantial increase over the
assumed LISM temperature of 10,900 K assumed for model 1. 
A further increase in the component 1 temperature occurs in the
downstream region.

\item[(v)] The number density of component 1 crossing the TS is $\sim
0.07$ cm${}^{-3}$.
This is approximately half 
the assumed incident LISM number density, an effect termed ``filtration''.
Between the TS and 10 AU from the Sun
in the upstream region, this density varies only weakly,
following a rough power law ($\sim R^{0.25}$, with $R$ the heliospheric
radius).
In the downstream direction, component 1 densities are lower within the
heliosphere and the gradient is 
somewhat steeper, with density increasing as $R^{0.35}$.

\item[(vi)] The upstream neutral gas is 
decelerated from $-26$ km/s in the LISM
to $-19$ km/s at the TS in the region of the nose.
Deflection of the flow also reduces the radial velocity component
at angles away from the nose.

\item[(vii)] Zank {\it et al.} (1996a) point out the possibility that the HP is
time dependent due to an inwardly directed ion-neutral drag term which provides
an effective ``gravitational'' term for a stratified fluid (which then
introduces the possibility of Rayleigh-Taylor-like instabilities). 
The time scale of
180 years and the $\sim 3$ AU amplitude of the oscillation suggest that this is
unlikely to be important. 

\end{description}

\subsection{One-Shock Model}

As described in the introduction,
since the SIC plasma thermal
sound speed is thought
to be $\sim$14 km/s, and the relative Sun-SIC
velocity is $\sim$26 km/s, a two-shock heliosphere is often
assumed to be necessary (Baranov {\it et al.} 1971).  
However, this neglects the effects of restoring forces
due to magnetic or cosmic-ray pressure, which could
enhance the effective sound speed.
Unfortunately, 
current knowledge of either the local interstellar magnetic field or
cosmic ray pressure is rudimentary.  
The nominal interstellar
pressure contributed by cosmic rays
(Ip \& Axford 1985) is $\sim 10^{-12}$ dyne cm$^{-2}$ 
(which gives $p/k \sim$7200 cm$^{-3}$ K) 
with perhaps $\sim (3 \pm 2) \times 10^{-13}$ dyne cm$^{-2}$
contributed by cosmic rays of energy 
less than about 300 MeV per nucleon, which may be expected to
couple to the plasma on heliospheric scales.
Unfortunately, these estimates are uncertain, particularly at MeV energies. 
But it is important to note that a cosmic ray
pressure of $3 \times 10^{-13}$ dyne cm$^{-2}$, combined with
the interstellar plasma thermal pressure from Table 1 
($2 \times 10^{-13}$ dyne cm$^{-2}$)
yields a total pressure of 
$5 \times 10^{-13}$ dyne cm$^{-2}$ (or $p/k \sim$3600 cm$^{-3}$ K).
This would be sufficient to increase the LISM sound speed 
to $\sim 27$ km/s and force the LISM inflow to be barely subsonic. 
Alternatively, a magnetic field strength of 3 $\mu$G  
with n$_{e}$=0.05 cm$^{-3}$ gives an Alfven speed of $v_a$=29 km/s,
which would enhance the magnetosonic speed
and also invalidate a two-shock model.

In view of the comments above, we also consider models with
subsonic LISM flow, which will not have a bow shock and
therefore may resemble the Parker (1963) model. 
For model 2, we again take n$_{e}$=0.07 cm$^{-3}$, n$_{H}$=0.14 cm$^{-3}$, 
and T=10,900 K (see Table 1) for consistency with model 1, but a larger 
``effective'' temperature is used in
determining the pressure, to account for the added contribution
from cosmic rays (and perhaps the magnetic field).
This effective temperature is defined in terms of a parameter $\alpha$,
as defined above.
The value $\alpha = 9.1$, shown in Table 1, has been
chosen for model 2, 
so the upstream Mach number is reduced from 1.5 
(as in model 1) to 0.7.
No charge exchange is assumed to occur between the
cosmic rays and neutrals due to the former's low number density.

Plots of the one-shock simulations are presented in 
Figure 2, 
where (b)
illustrates the 2D distribution of component 1 neutrals in the same format as 
used in Figure 2a.  
Figure 2c-e shows $\alpha$ Cen line-of-sight 
profiles for the density, velocity and temperature of component 1, using
dotted curves for model 2.   
Again, we merely summarize the main features here
(and see Zank {\it et al.} 1996a for further details).

\begin{description}

\item[{\bf (i)}] Although a bow shock is absent, some adiabatic compression of
the incident interstellar flow is evident. 
This gradual compression forms a lower amplitude
hydrogen wall that is more extended in the radial direction.
It is also less extended in the tangential direction because of the
localized nature of the adiabatic compression.   
The density of the wall in the upstream direction is
only $\sim 0.21$ cm$^{-3}$ (though still larger than the incident 
LISM n$_{H}$=0.14 cm$^{-3}$).
However, because it is wider, its column
density is comparable to the two-shock case.

\item[{\bf (ii)}] The heliosphere is less distorted along the axis of symmetry
than for the two-shock case,
and is smaller due to the higher assumed LISM pressure.

\item[{\bf (iii)}] In the vicinity of the nose, the number density of 
component 1 
flowing across the TS is $\sim
0.06$ cm${}^{-3}$ with a velocity of $\sim -20$ km/s, 
almost identical to the two-shock model. 

\item[{\bf (iv)}] Since the H wall has a smaller transverse extent
than the two-shock model, it
is less pronounced  along the sidestream sightline. 
This may allow the one- and two-shcok models to
be observably different not only upstream, but sidestream as well. 

\item[{\bf (v)}] The upstream and downstream temperature characteristics of the
heliospheric component 1 differ significantly between the one- and two-shock
models. 
In the upstream direction of the one-shock model, $\sim$2,000 K of 
cooling for the neutrals is predicted. 
A temperature asymmetry between upstream and downstream heliospheric neutrals
is again present, but the downstream temperatures are 
markedly lower than predicted by the two shock model.

\end{description}

\subsection{Intermediate Case--- Barely Subsonic}

To bridge the gap between models 1 and 2, we
have computed a third model in which the inflow Mach number
was chosen to take the intermediate value 0.9 (model 3, Table 1).
In the interest of increased realism,
at the cost of sacrificing consistency with models 1 and 2,
we reduced the inflow temperature to the
more realistic value $T=7,600$ K, and to compensate we increased
the plasma density slightly to
n$_{e}$=0.1 cm$^{-3}$ so as to preserve the incident plasma heat flux.
Further variations in the densities would be required to span the
observably allowable domain, but this is left for future
work, the present paper being restricted
to the three models in Table 1.

The model 3 results for the component 1
neutrals are depicted by dot-dashed curves in Figure 2c-e.
The overall structure and distribution are similar to model 2, 
underscoring the qualitative connection between subsonic models.
The quantitative attributes of the
hydrogen wall are generally intermediate to models 1 and 2,
presumably owing to the intermediate value of the Mach number,
which appears to be the most important single parameter.

\section*{References}
Baranov, V. B. \& Malama, Y. G. 1993, 
{\em J. Geophys. Res., 98}, 15, 157  \\
Baranov, V. B. \& Malama Y. G. 1995, 
{\em J. Geophys. Res., 100}, 14, 755  \\
Baranov, V. B., Krasnobaev, K., \& Kulikovsky, A.
1971, {\em Sov.\ Phys.\ Dokl.\ 
15}, 791 \\
Bertin, P., Lallement, R., Chassefiere, E., \& Scott, N. 1993, JGR, 98, 15193 \\
Bertin, P., Vidal-Madjar, A., Lallement, R., Ferlet, R.,
\& Lemoine, M. 1995, {\em Astron.\ Astrophys., 302}, 889  \\
Brekke, P., Kjeldseth-Moe, O. Bartoe, J.-D. F., Brueckner, G., 
\& Van Hoosier, M. E. 1991, ApJS, 75, 1337 \\
Dring, A. R., Murthy, J., Henry, R. C., Landsman, W., Audouze, J.,, Brown, A., 
Linsky, J. L., Moos, W., \& Vidal-Madjar, A. 1997, ApJ, submitted \\
Frisch, P.C. 1994, 
{\em Science, 265}, 1423 \\
Frisch, P. C.  1993, \apj, 407, 198 \\
Frisch, P. C., Welty, D. E., Pauls, H. L., Williams, L. L., \& Zank G. P. 
1996, BAAS, 28, 760  \\
Frisch, P. C. 1996, Space Sci.\ Rev., in press \\
Gloeckler, G. 1996, Space Sci.\ Rev., in press \\ 
Hall, D. T. 1992, 
Ph.D.\ thesis, Univ.\ of Arizona, Tucson \\
Holzer, T.E. 1972, 
{\em J.\ Geophys.\ Res.}, 77, 5407 \\
Holzer, T.E. 1979, in Solar System Plasma Physics, ed.'s C.F.\ Kennel, L.J.\
Lanzerotti, and E.N.\ Parker, North Holland, p. 103 \\
Landsman, W. B., Henry, R. C., Moos, H. W., \& Linsky, J. L. 1984, \apj,
285, 801 \\
Landsman, W. B., Murthy, J., Henry, R. C., Moos, 
H. W., Linsky, J. L., \& Russell, J. L. 1986, 303, 791 \\
Ip, W.-H. \& Axford, W. I. 1985, 
{\em Astron.\ Astrophys.}, 149, 7 \\
Lallement, R., Ferlet, R., Lagrange, 
A. M., Lemoine, M., \& Vidal-Madjar, A.  1995,  
{\em Astron.\ Astrophys.}, 304, 461  \\
Linsky, J. L., Brown, A., Gayley, K. G., Diplas, A., Savage, B. D.,
Ayres, T. R., Landsman, W., Shore, S. N., \& Heap, S. R. 1993,
ApJ, 402, 694. \\
Linsky, J. L. \& Wood, B. E. 1996, ApJ, 463, 254 \\
Mullan. D. J., \& Arge, N. 1995, J.\ Geophys.\ Res., 101, 2535 \\
Parker, E. N. 1963, 
{\em Interplanetary Dynamical Processes}, Interscience, New York \\
Pauls, H. L. \& Zank, G. P. 1996a, 
{\em J.\ Geophys.\ Res.}, 101, 17,081 \\
Pauls, H. L., \& Zank, G. P. 1996b, 
{\em J.\ Geophys.\ Res.}, submitted \\
Pauls, H. L., Zank, G. P., \& Williams, L. L. 1995, 
{\em J.\ Geophys.\ Res.}, 100, 21,595 \\
Piskunov, N., Wood, B. E., Linsky, J. L., Dempsey, R. C., \& 
Ayres, T. R. 1997, \apj, 474, 315 \\
Ripkin, H. W., \& Fahr, H. J. 1983, 
{\em Astron.\ Astrophys.}, 122, 181 \\
Slavin, J.  1989, \apj, 346, 718 \\
Williams, L.L., Hall, D. T., Pauls, H. L., \& Zank, G. P. 1996b, 
{\em Astrophys.\ J.,} in press\\
Williams, L. L., Pauls, H. L., Zank, G. P., \& Hall,
D. T. 1996a,
in {\em Proc.\ Solar Wind 8}, Am.\ Inst.\ of
Phys., Conf. Proc. 382, D. Winterhalter {\it et al.}, eds., p. 609-612 \\
Witte, M., Rosenbauer, H., Banaszkiewicz, M., \& Fahr, H. 1993, AdSpR, 
13, 121 \\
Wood, B. E., Alexander, W. R., \& Linsky, J. L. 1996, ApJ, in press \\
Zank, G. P. \& Pauls, H. L. 1996, {Space Sci.\ Rev.}, 78, 95 \\
Zank, G. P., Pauls, H. L., Williams, L. L., \& Hall, D. T.
1996a, {J.\ Geophys.\ Res.}, 101, 21,639 \\
Zank, G. P., Pauls, H. L., Williams, L. L., \& Hall, D. T. 1996b,
in {\em Proc.\ Solar Wind 8}, Am.\ Inst.\ of
Phys., Conf. Proc. 382, D. Winterhalter {\it et al.}, eds., p. 654-657 \\

\newpage

\begin{tabular}{p{0.5in}
p{0.7in}p{0.75in}
p{0.75in}p{0.7in}p{0.75in}p{0.55in}}
\multicolumn{7}{c}{Table 1.  Model Inflow Parameters} \\
\hline
Model & Mach \# & $n_e$ (cm$^{-2}$)  & $n_H$ 
(cm$^{-2}$) & $T$ (K) & $v$ (km/s) & $\alpha$ \\
\hline
1 & 1.5 & 0.07 & 0.14 & 10900 & -26 & 2 \\
2 & 0.7 & 0.07 & 0.14 & 10900 & -26 & 9.1 \\
3 & 0.9 & 0.1 & 0.14 & 7600 & -26 & 7.9 \\
\hline
\end{tabular}

\newpage
\vspace{0.5in}
\begin{tabular}{p{0.4in}p{0.6in}
p{0.8in}p{0.6in}p{0.6in}p{0.6in}p{0.7in}p{0.5in}}
\multicolumn{8}{c}{Table 2.  Average Properties of Model Hydrogen Walls} \\
\hline
Model & $T_{hw}$ & 
$v_{hw}$ & 
$n_H$ & 
$\Delta L_{hw}$ &
$D_{hp}$ &
$N_{hw}$ &
$\tau_{0.1}$ \\
 & (K) & (km/s) & (cm$^{-3}$) & (AU) & (AU) & ($10^{14}$ cm$^{-2}$) &  \\
\hline
1 & 23000 & -4.3 & 0.20 & 210 & 160 & 6.7 & 2.8 \\
2 & 14700 & -4.3 & 0.19 & 125 & 140 & 3.8 & 0.6 \\
3 & 16000 & -4.1 & 0.12 & 325 & 220 & 6.3 & 1.2 \\
LW & 29500 & -8 & 18/$L_{hw}$ & 18/$n_H$ & -- & 3 & 1.1 \\
\hline
\end{tabular}

\newpage
\centerline{Figure Captions}
\noindent
{\bf Fig. 1}.
The solid curves
are GHRS L$\alpha$ profiles toward (a) $\alpha$ Cen A
and (b) $\alpha$ Cen B, from LW.
The upper dashed curve is the assumed intrinsic stellar L$\alpha$
emission profile.  The dotted curve shows the intrinsic stellar emission line
after absorption by a purely LISM cloud with with $N_H = 4.5 \times
10^{17}$ cm$^{-2}$, $b = 9.3$ km s$^{-1}$, and
$v = -18.2$ km s$^{-1}$.

\noindent
{\bf Fig. 2}.
Results are depicted for the neutrals of
interstellar origin from the 2D heliospheric models in Table 1.
The contours denote log $T$, the arrows indicate the flow direction,
and the shading gives the density normalized to the inflowing
interstellar gas density.
The results  
from model 1 are in (a), and from model 2 in (b), with model 3 results
being intermediate to these. 
The velocity, temperature, and density profiles along the $\alpha$
Cen sightline for all three models are shown in (c-e), where
the dashed curve is model 1 (supersonic), dotted is model 2 (subsonic),
and dot-dashed is model 3 (barely subsonic).
Negative velocities imply flow toward the Sun.

\noindent
{\bf Fig. 3}.
Similar to Figure 1, except that absorption from the three heliospheric models
is included.
All curves from Figure 1 are reproduced as solid lines, while the dashed curve
is for model 1 (M=1.5), dotted for model 2 (M=0.7), and dot-dashed for 
model 3 (M=0.9).  The profile in (a) depicts the heliospheric absorption
for the $\alpha$ Cen A profile, whereas (b) is for $\alpha$ Cen B.
The red edge of the LISM absorption feature is best fit by model 3, and
note that none of the models can fit the blue edge.
In order to suggest the possible importance of the $\alpha$ Cen heliosphere
in fitting the blue edge of the LISM feature,
profiles (c) and (d) depict the absorption in the $\alpha$
Cen A and B profiles (respectively) that would result if identical
heliosphere models also surrounded $\alpha$ Cen.

\noindent
{\bf Fig. 4}.
Contribution functions to the optical depth in the heliospheric model
at 0.1 {\AA} from line center (eq. 1), plotted as a function of the distance from the Sun.
The axes are scaled such that the area under the curves gives
the total $\tau_{hw}(0.1)$.
The dashed curve is model 1, dotted model 2, and dot-dashed
model 3, for comparison to Figure 2.
Comparison with Figure 2e shows the connection to the elevated temperature
structure.

\newpage
\noindent
{\bf Fig. 5}.
Contour plots of the equivalent width (in \AA) of heliospheric L$\alpha$
absorption as a function of LISM H I column depth ($N_H$) and line of sight.
We assume a LISM model with $b=9.3$ km s$^{-1}$, $v=-29$ cos $\theta$ km/s.
The variable $N_H$ is given by
the distance to the origin, in the scale marked on
both the abscissa and ordinate, as in a polar plot.
The angle $\theta$ in the polar plot
is the angle between the nose direction and 
the sightline to potential stellar targets.
The sightlines to 36 Oph, $\alpha$ Cen, and 31 Com,
at $\theta = 12\deg, 52\deg$, and $72\deg$ respectively,
are indicated by dotted lines.
Model 1 results (supersonic inflow) 
are shown in (a), and model 2 results (subsonic inflow) are in (b).
Model 3 is intermediate to these.
The GHRS Echelle A can easily resolve a total absorption corresponding
to the 0.008 \AA \ contour.



\noindent
{\bf Fig. 6}.
A synthetic profile consisting of the LISM and two 
conductive interfaces (dot-dash line) as predicted by Slavin (1989).
The assumed parameters are described in the text.
The dashed curve shows the absorption by the LISM alone.
The total amount of absorption seen is nearly
in agreement with the observations, except for a
telling lack of appropriate overall redshift
and an excess of highly broadened (high $T$) absorption.

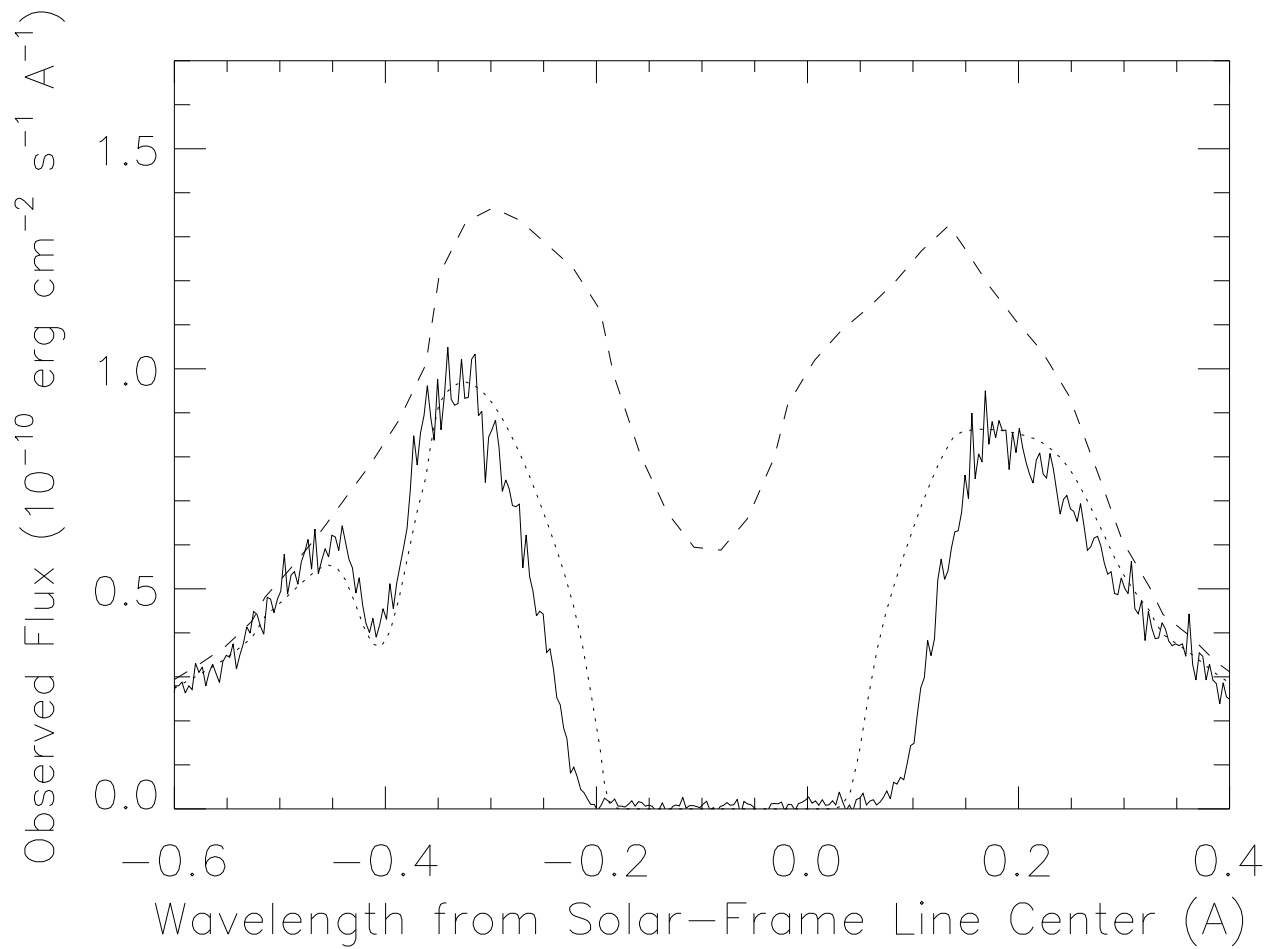
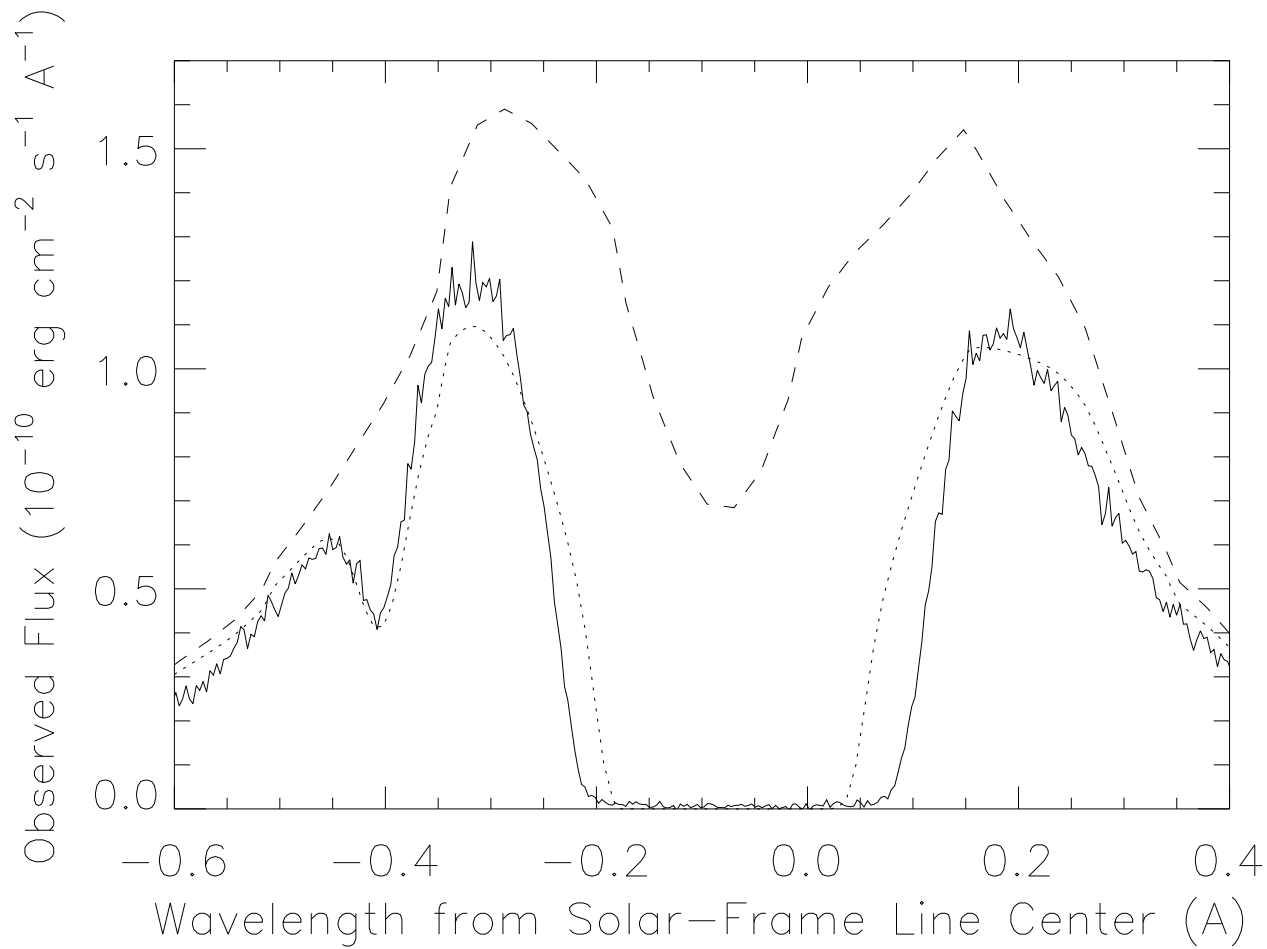
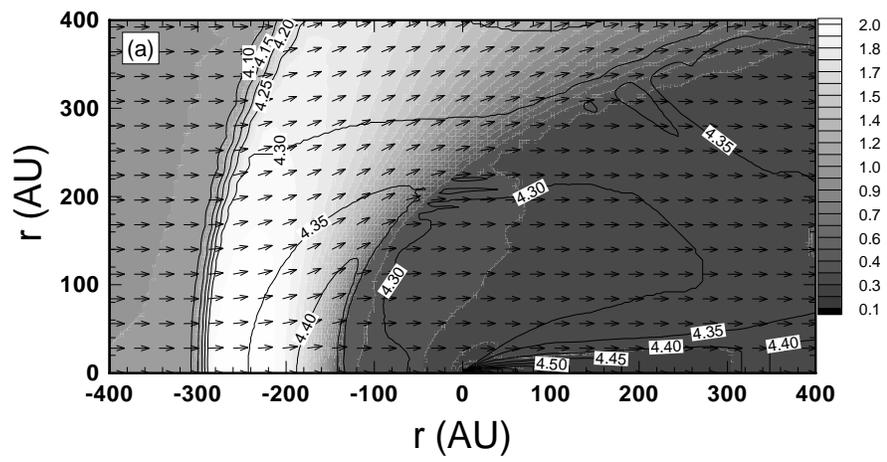
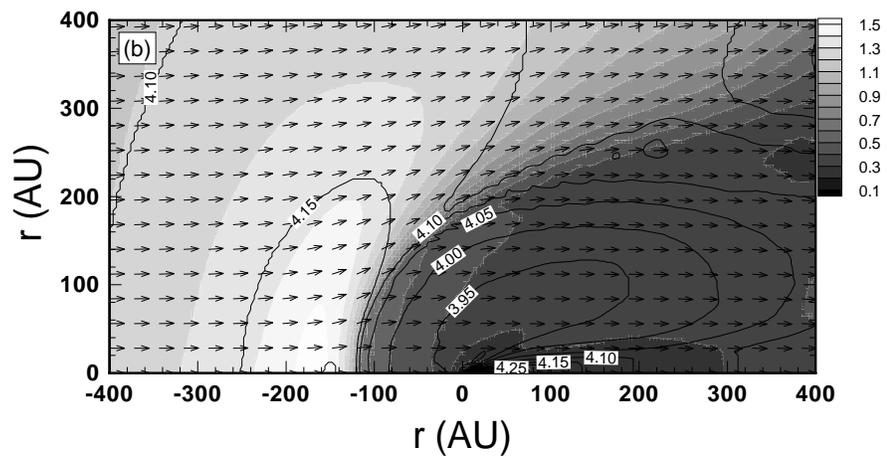
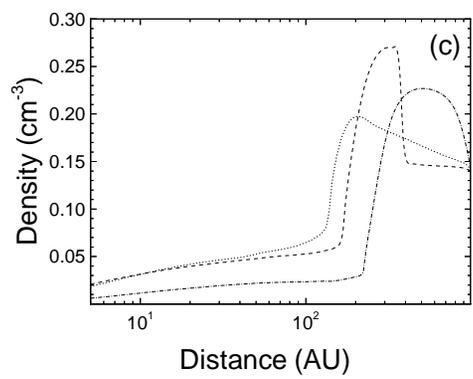
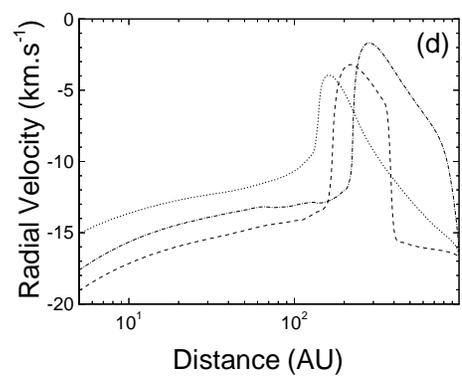
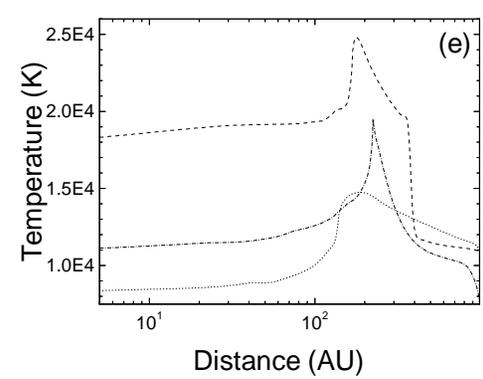
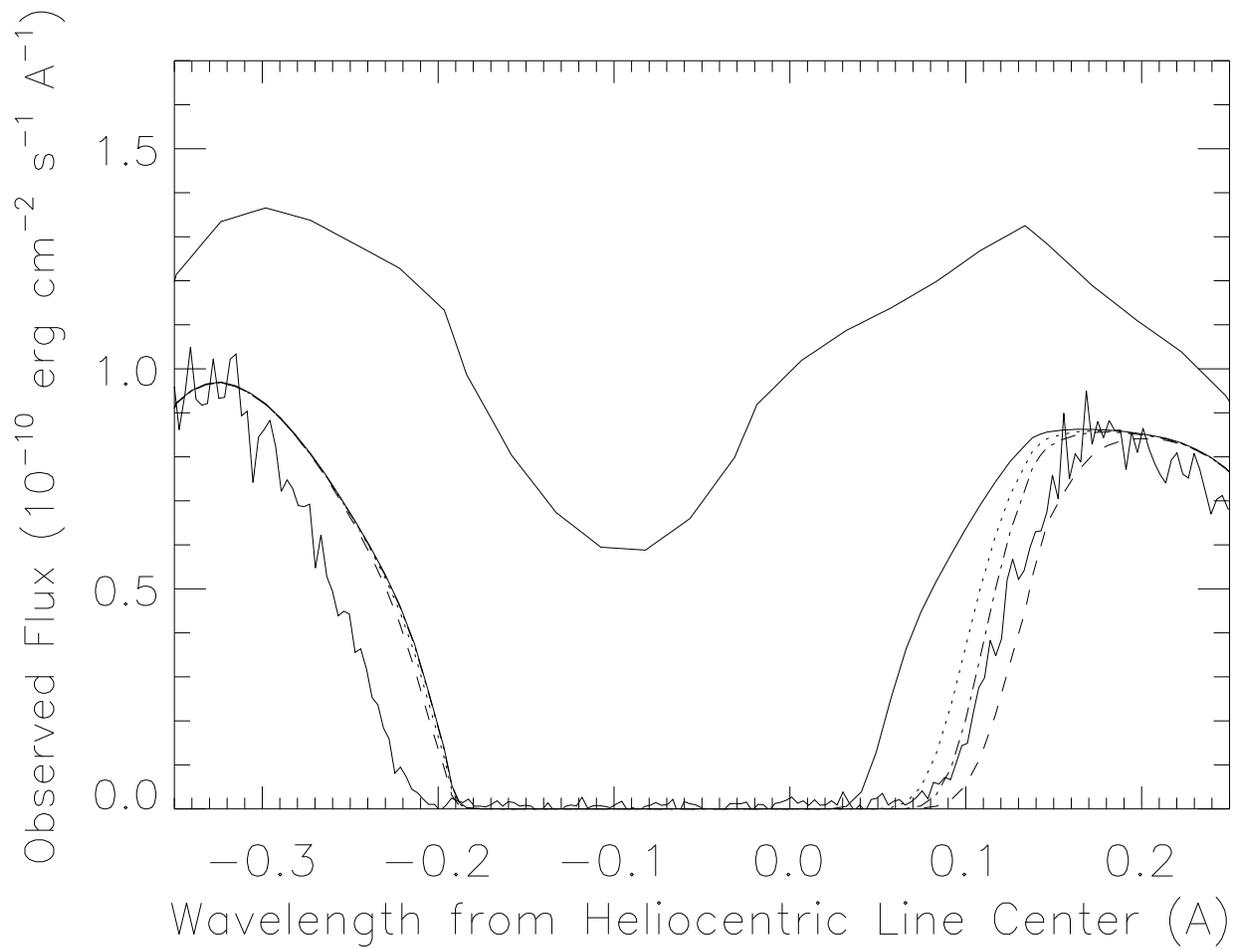
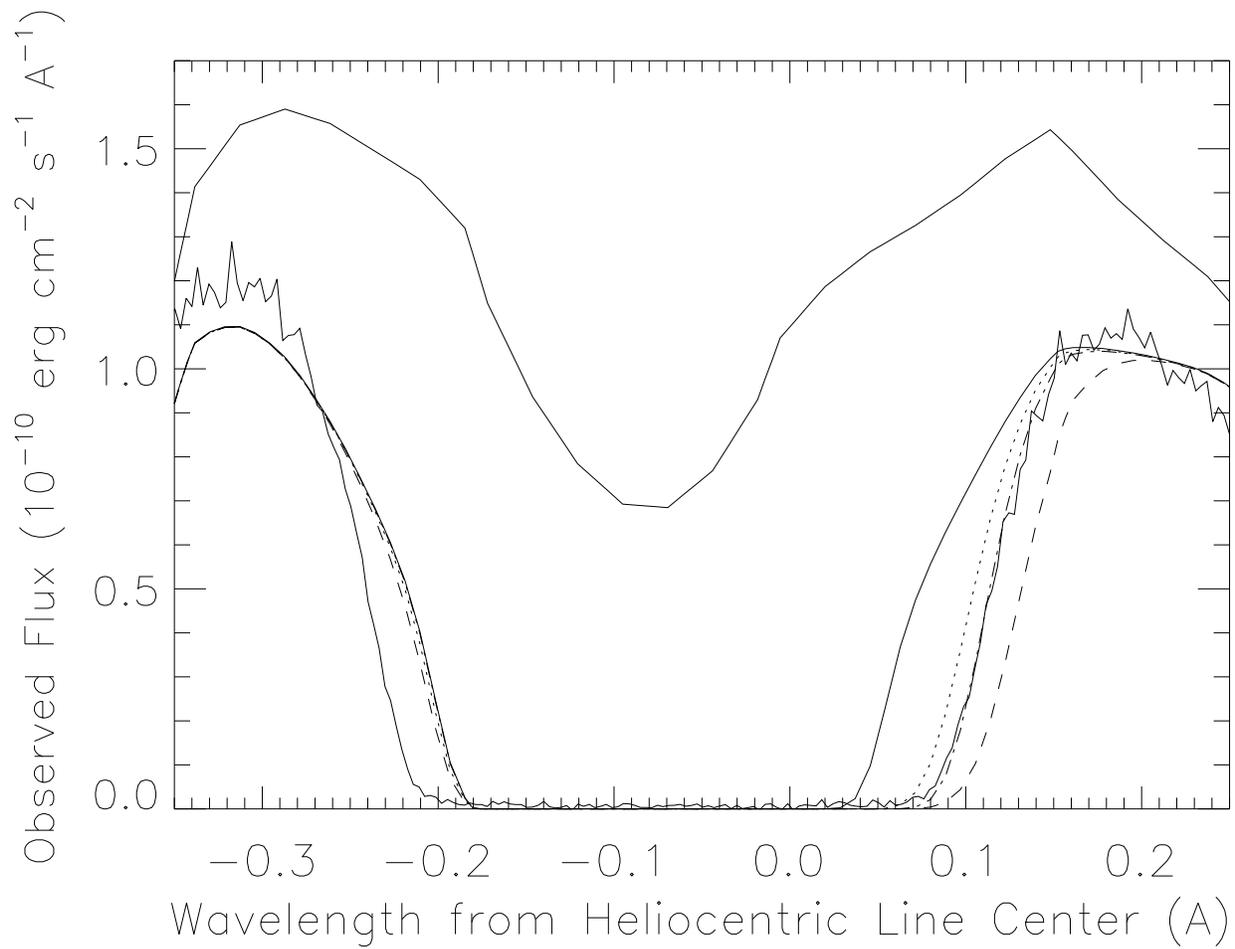
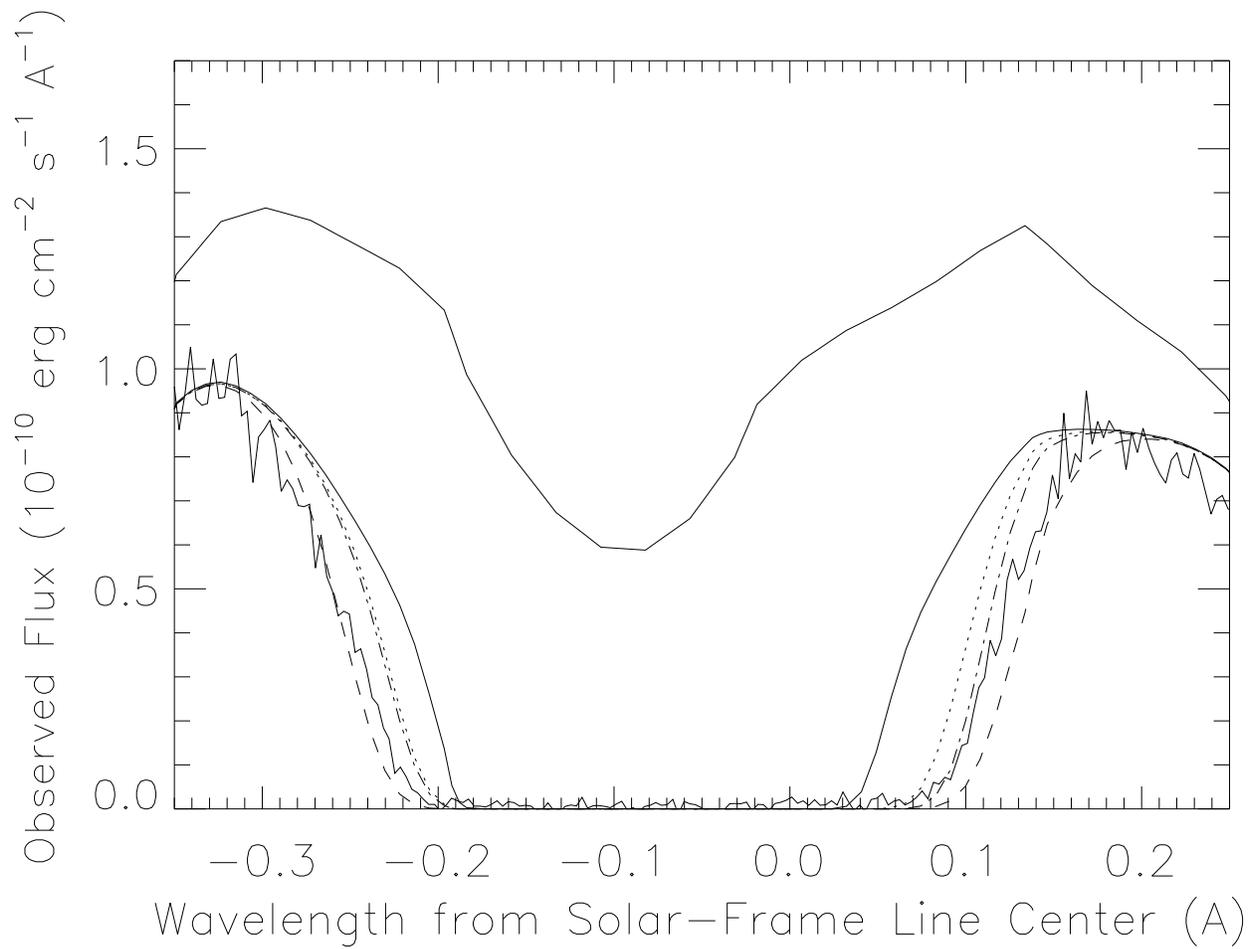
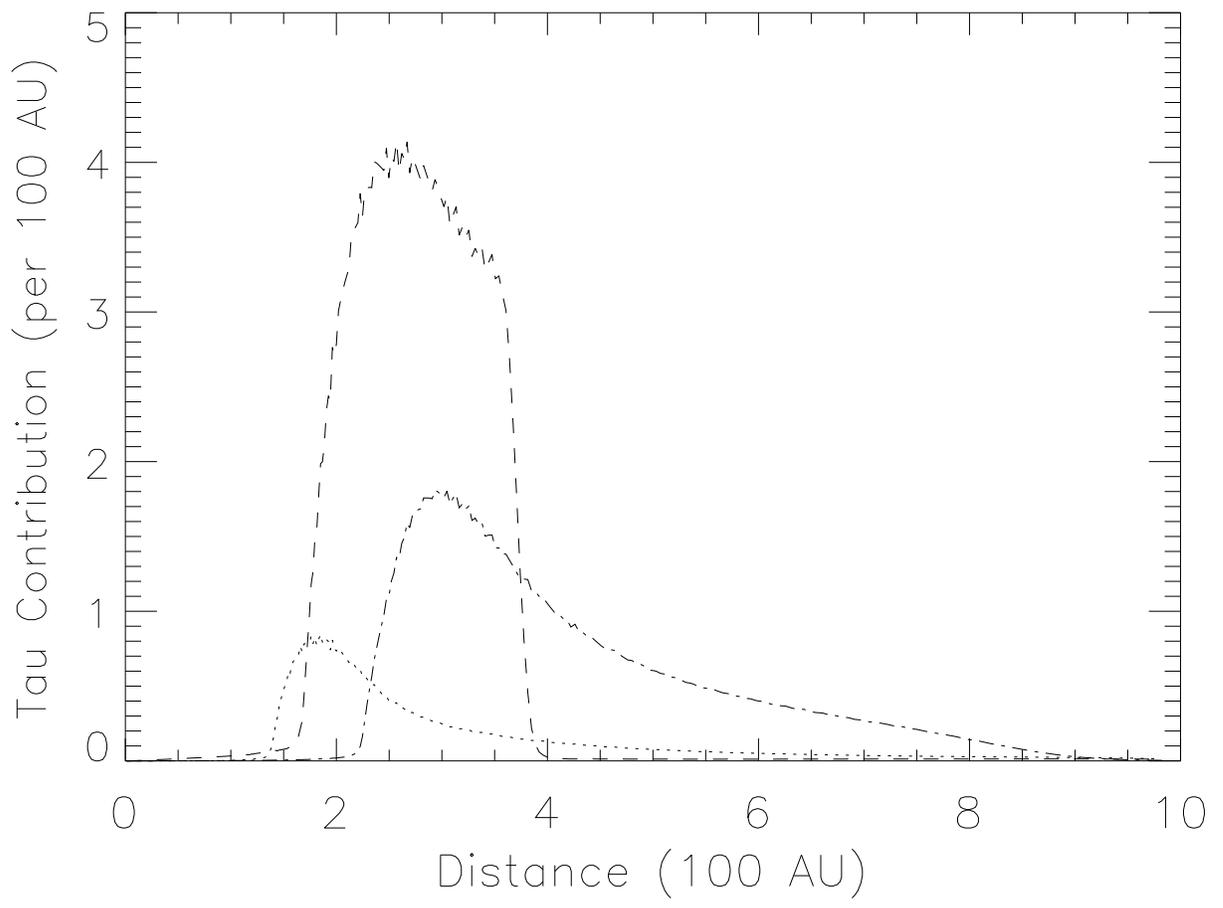
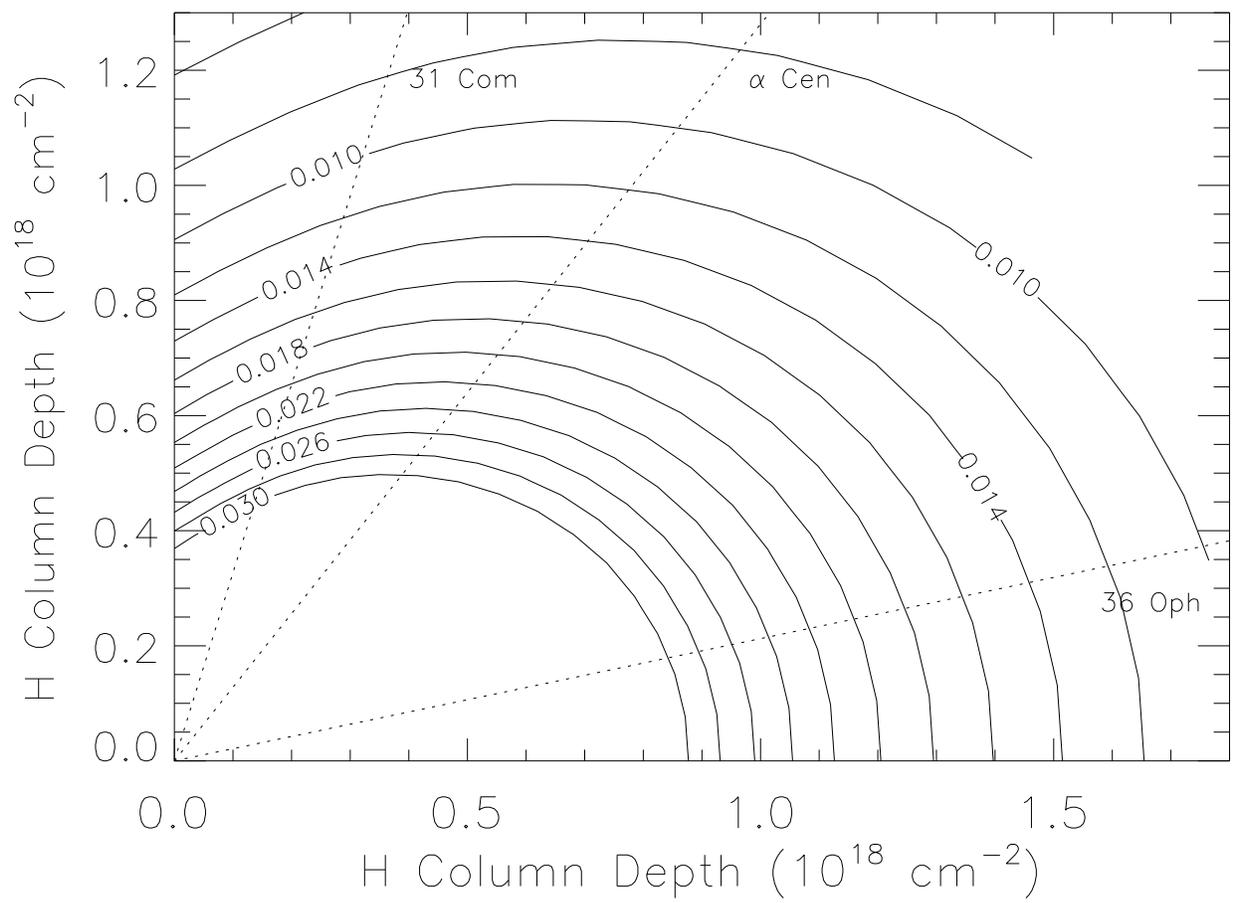
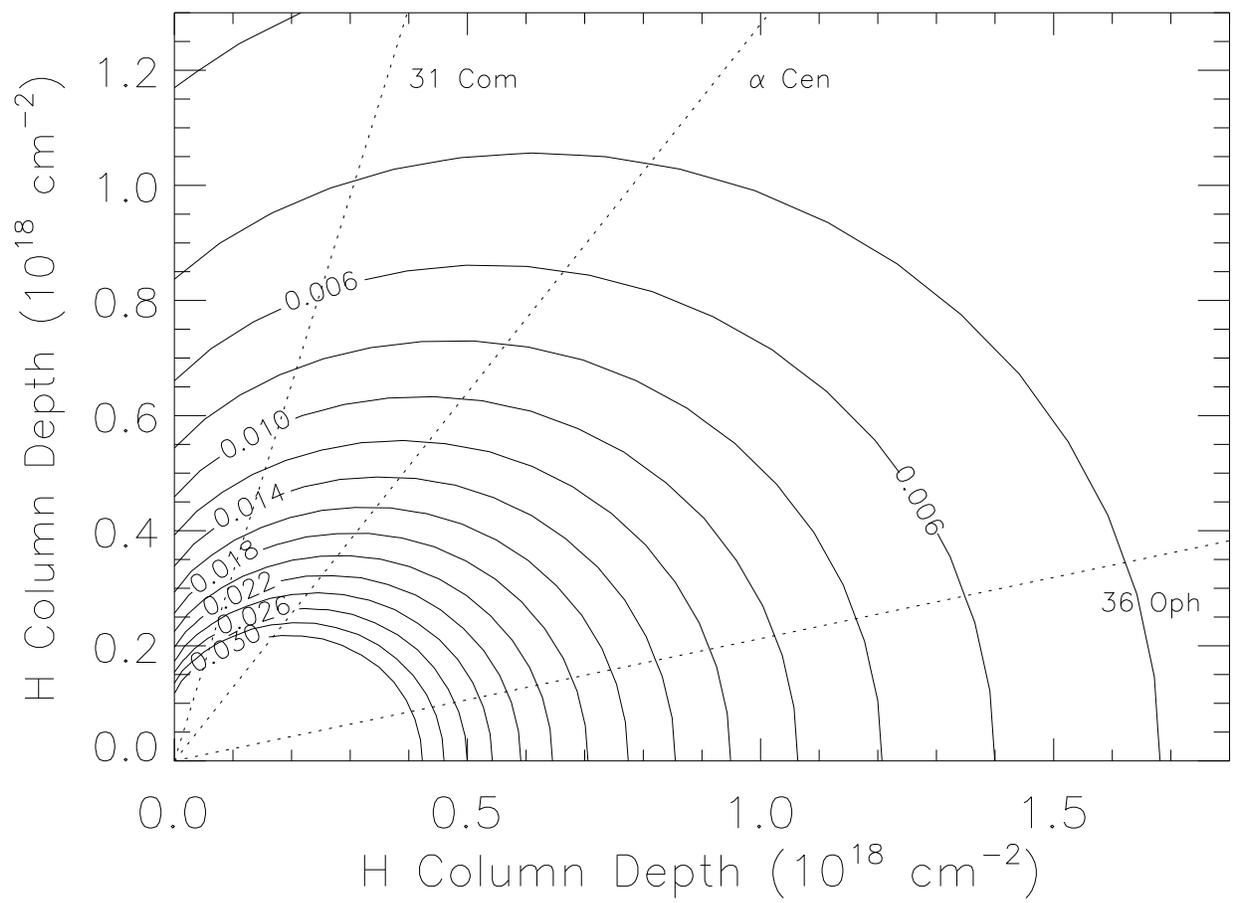
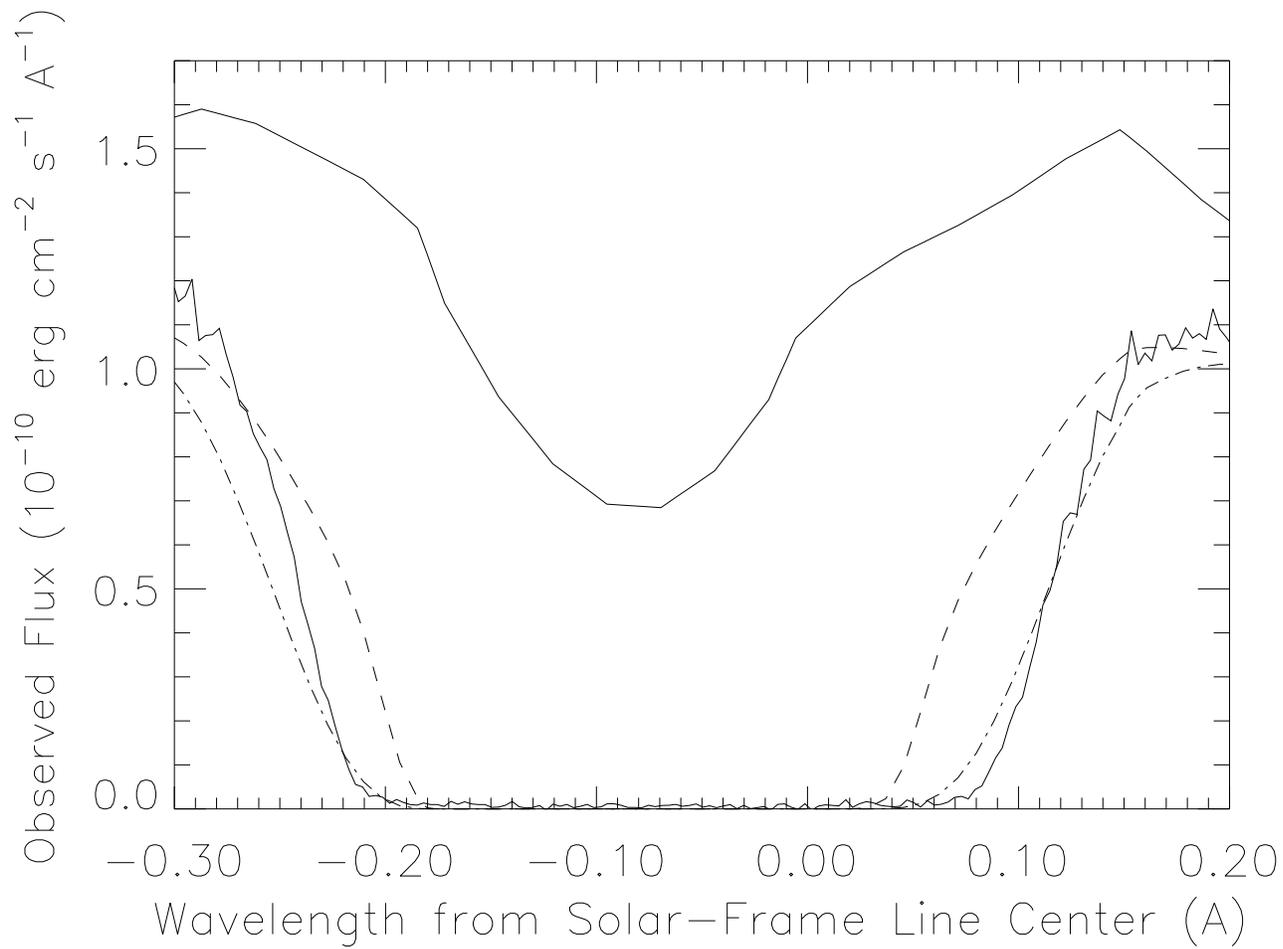
\end{document}